\documentclass[showpacs,showkeys,preprintnumbers,amsmath,amssymb,twocolumn,pre,floatfix]
 {revtex4}

\usepackage{amssymb,amsfonts}
\usepackage{amsmath}
\usepackage{hyperref}
\usepackage{graphicx}
\usepackage[usenames,dvipsnames]{xcolor}
\graphicspath{{Figures/}}
\begin{document}

\title{Nonlinear Dynamics of Dipoles in Microtubules:  Pseudo-Spin Model}

\author{Alexander I Nesterov}%
  \email{nesterov@cencar.udg.mx}
\affiliation{Departamento de F{\'\i}sica, CUCEI, Universidad de Guadalajara,
Av. Revoluci\'on 1500, Guadalajara, CP 44420, Jalisco, M\'exico}

\author{M\'onica F Ram\'irez}
 \email{monica.felipa@gmail.com}
\affiliation{ Departamento de F{\'\i}sica, CUCEI, Universidad de 
Guadalajara,
Av. Revoluci\'on 1500, Guadalajara, CP 44420, Jalisco, M\'exico}

\author{Gennady P  Berman}
 \email{gpb@lanl.gov}
\affiliation{Theoretical Division, T-4, Los Alamos National Laboratory, and 
the New
Mexico Consortium,  Los Alamos, NM 87544, USA}

\author{ Nick E Mavromatos}
 \email{nikolaos.mavromatos@kcl.ac.uk}
\affiliation{ King's College London, Physics Department, King's College London, Strand, London 
WC2R 2LS, UK}
 
\date{\today}

\begin{abstract}
We perform a theoretical study of the dynamics of the electric field excitations in a microtubule by taking into consideration the realistic cylindrical geometry, dipole-dipole interactions of the tubulin-based protein heterodimers, the radial electric field produced
by the solvent, and a possible degeneracy of energy states of  individual heterodimers. The consideration is done in the frames of the classical pseudo-spin model. We derive the system of nonlinear dynamical ordinary differential equations of motion for interacting dipoles, and the continuum version of these equations. We obtain the solutions of these equations in the form of snoidal waves, solitons, kinks, and localized spikes. Our results will help to a better understanding of the
functional properties of microtubules including the motor protein dynamics and the information transfer processes. Our considerations are based on classical dynamics. Some speculations on the role of possible quantum effects are also made.
\end{abstract}

\pacs{87.15.ht, 05.60.Gg, 82.39.Jn }

\keywords{Microtubules, dipole-dipole interaction}

\preprint{LA-UR-16-22690}

\maketitle

\section{Introduction}

Microtubules (MTs) are cylindrically shaped cytoskeletal biopolymers. They are found in eukaryotic cells and are formed by the polymerization of heterodimers built of two globular proteins, alpha and beta tubulin \cite{Amos}. The MTs can grow up to 50 $\mu m$ long (with an average length of 25 $\mu m$). The MTs are highly dynamic. In the growing phase, alpha and beta tubulins spontaneously bind one another to form a functional subunit that is called a heterodimer. In the shortening phase, the MT shrinks its length. A single MT can also oscillate between growing and shortening phases. The MTs perform many functions within the cell. In particular, the MTs support the cytoskeleton,  participate in the intracellular transport, provide the transportation of secretory vesicles, organelles, and intracellular substances, are  involved in cell division, and are believed to participate in the classical and quantum information transfer processes. 

Because a single MT is built of a set of macroscopic dipoles, the static and dynamic electric fields, generated by these dipoles, are crucial for understanding the functional properties of a single MT and the interactions between the MTs.

In \cite{Sataric1}, a classical one-dimensional  model of interacting dipoles with local $\phi^2-\phi^4$ potential and in the presence of a static electric field is introduced, for describing the energy-transfer by kinklike excitations in cell MTs, in terms of a single variable (elastic degree of freedom).  A similar model was used in \cite{Sataric2} to study the influence of d.c. and a.c. electric fields on the dynamics of MTs in living cells. In \cite{Sataric3,Sataric4,Sataric5} the extension of the model considered in \cite{Sataric1,Sataric2} was proposed in order to elucidate the unidirectional transport of cargo via motor proteins such as kinesin and dynein, and for describing the  nonlinear dynamics within a MT and solitonic ionic waves along the microtubule axis. In \cite{Slyadnikov1}, the physics of the dipole system of a neuron cytoskeleton MT is discussed, based of the quantum approach, where the tunneling effects on individuals heterodimers are taken into account. The possible effects of quantum coherence and entanglement in brain MTs and efficient energy and information transport were studied in \cite{Nick11,Nick12,Nick13,Nick21,Nick22}, where it was argued that under certain circumstances, in particular in the case of {\it in vivo} MT, quantum coherence may be maintained up to micro seconds before collapsing in a classical state. This should be sufficient for `quantum wiring' of the MT system, in analogy with recently claimed long-lasting (femtoseconds) quantum correlation effects in algae~\cite{algae}. From a theoretical point of view, quantum corrections to the classical solitonic states (obtained as a solution of the dynamical system of equations of MT models, as done in the present article) have also been considered in a WKB approximation in \cite{Nick11,Nick12,Nick13}. 
The dielectric measurements of individual MTs using the
electroorientation method are described in \cite{Minoura}. The 
multi-level memory-switching properties of a single brain MT were studied experimentally in \cite{Sahu}. 

In spite of the many models of the MTs introduced and studied in the literature, no consensus is reached on the relations between the outcomes of these models and the MT functionality. 

In this paper, we introduce and study theoretically a generalized model of a single MT which takes into account the realistic cylindrical geometry of the MT, the dipole-dipole interactions of the tubulin-based protein heterodimers, the radial electric field produced by the solvent, and a possible degeneracy of the energy states of the individual heterodimers. Our consideration is done in the framework of the classical ``pseudo-spin" model, as the length of the individual dipole of the heterodimer is assumed to be constant. 

We derive the system of nonlinear dynamical partial differential equations of motion for interacting dipoles of the heterodimers, and the continuum version of these equations. We obtain the partial solutions of these equations in the form of snoidal waves, solitons,  kinks, and localized spikes, and describe their properties. We hope that our results will help to understand better  the relations between the electric excitations and the functional properties of the MTs such as motor protein dynamics and the information transfer processes.

The structure of the paper is the following. In Section II, we describe our model. In Section III, we apply our approach to analyze the dynamics of the system, and present the results of the numerical simulations for both exact and approximate solutions. In the Conclusions section we summarize our results and formulate some challenges for future research. 

\section{Description of the model}

\begin{figure}[tbh]
\begin{center}
 	\scalebox{0.5}{\includegraphics{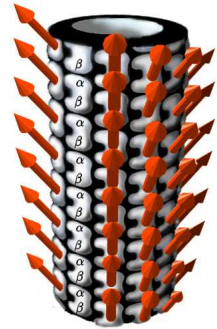}}\\
  \end{center}
 \caption{The structure of the cytoskeleton microtubule.  The arrows indicate the orientation of the permanent dipole moments of the  tubulin heterodimers with respect to the surface of a microtubule.
  \label{MT}}
  \end{figure}

 MTs are realized as hollow cylinders typically formed by $13$ parallel protofilaments (PFs) covering the wall of MT \cite{TBHM,Sataric3}. The outer diameter of a MT is about 25 nm, and the inner diameter is about 15 nm.  Each PF is formed by $(\alpha,\beta)$-tubulin heterodimers (Fig. \ref{MT}). 
Due to their interaction with the complex biological environment (solvent) the MTs may experience a  strong radial electrostatic field leading to the additional (radial) polarization of tubulins  \cite{BSJ}. 

The tubulin heterodimer contains approximately 900 amino acid residues with the number of atoms about 14000. The total mass of the heterodimer can be estimated as, ($M \approx 1.84\cdot 10^{-19} \rm g $). Each heterodimer can be considered as effective electric dipole with $\alpha$ and $\beta$ tubulin being as positive and negative side of dipole, respectively \cite{SM1}. (See Fig. \ref{MT}.)

We treat each dipole as a classical pseudo-spin, $\mathbf S_i$, with a constant modulus.  The potential energy of the system can be written as:
\begin{align}
{ U}_0=&\sum_{\langle  i,j \rangle}    J_{ij}\big( {\mathbf S}_i 
\cdot{\mathbf S}_j  - 3 (\mathbf S_i\cdot {\mathbf e}_{ij} )(\mathbf 
S_j\cdot {\mathbf e}_{ij}) \big) \nonumber \\
& -  B \sum_{i}\mathbf S_i\cdot {\mathbf e}_r,  
\label{H2}
\end{align}
where $\mathbf e_{ij}$ is a unit vector  parallel to the line connecting  the dipoles, ${\mathbf S}_i $ and ${\mathbf S}_j$. The  scalar product is understood as follows: ${\mathbf S}_i 
\cdot{\mathbf S}_j ={S}^1_i { S}^1_j + {S}^2_i { S}^2_j + {S}^3_i { S}^3_j$. 
The first sum describes  the dipole-dipole interaction, and the second one characterizes the effect of the transversal (radial) electrostatic field acting on the dipoles.

Since the MTs may exhibit ferroelectric properties at room
temperature, one can consider the MT as a ferroelectric system \cite{Sataric3,TBCC}. To include into consideration the ferroelectric properties of the MT, we adopt the approach developed in \cite{THST}. In this case, the overall effect of the environment on the effective spin, $\mathbf S_i$, is described by the  double-well quartic on-site potential,
\begin{align}
V({\mathbf S}_i ) =  P ({\mathbf S}_i \cdot  {\mathbf e}_z	)^2 + Q ({\mathbf S}_i\cdot  {\mathbf e}_z	)^4.
\end{align}

It is convenient to parameterize  the pseudo-spin $\mathbf S_i$ by  the unit vector $\mathbf n_i$, as: $\mathbf S_i = S \mathbf n_i$. Then, the total potential energy of the system can be written as,
\begin{align}
&{U}=S^2\sum_{ \langle  i,j \rangle}    J_{ij}\big( {\mathbf n}_i 
\cdot{\mathbf n}_j  - 3 (\mathbf n_i\cdot {\mathbf e}_{ij} )(\mathbf 
n_j\cdot {\mathbf e}_{ij}) \big)  \nonumber \\
&+ \sum_{i} \big(PS^2 ({\mathbf n}_i \cdot  {\mathbf e}_z	)^2 + QS^4 ({\mathbf 
n}_i\cdot  {\mathbf e}_z	)^4 -
B S \mathbf n_i\cdot {\mathbf e}_r \big).
\label{H2a}
\end{align}
The dynamics of the system is described by the discrete 
Euler-Lagrange equations \cite{BYB}:
\begin{align}\label{EqM1a}
\frac{d\mathbf n_i}{d t} = \frac{1}{I}\mathbf L_i \times \mathbf n_{i}, \\
\frac{d\mathbf L_i}{d t} = \mathbf n_i \times \mathbf E_{i},
\label{EqM1b}
\end{align}
where $  \mathbf E_{i} =-{\partial U}/{\partial {\mathbf n}_i}$, and $\mathbf L_i  $ is the angular momentum of the dipole located at the site $i$, its moment of inertia being $I$. 
Substituting ${\mathbf L_i } =I {\mathbf n_i }\times \dot {\mathbf n}_i $ into Eq. (\ref{EqM1b}), we obtain
\begin{align}
	I\frac{d^2\mathbf n_i}{d t^2} + I\mathbf n_i \bigg(\frac{d\mathbf n_i}{d t}\bigg)^2 = \mathbf E_i - \mathbf n_i (\mathbf n_i \cdot \mathbf E_i ).
	\label{EQM1}
\end{align}

The equations of motion can be obtained from the classical action,
\begin{align}
	S=  \int { L_c} dt,
	\label{Scl1}
\end{align}
where $L_c = T-U + \Sigma _i \lambda_i ({\mathbf n}_i \cdot  {\mathbf n}_i -1)$.

 The kinetic energy of the system is,
\begin{align}
	T={\Sigma _i} \frac{\mathbf L^2_i }{2 I} = {\Sigma _i} I\frac{\dot 
	{\mathbf 
	n}^2_i }{2 }  ,
	\end{align}
 and the Lagrange multiplier, $\lambda_i$, provides the constraint, 
 $\mathbf 
 n_i \cdot \mathbf n_i =1$, to be satisfied.

The Euler-Lagrange equations, following from the variation of the action, $\delta S=0$, take the form,
\begin{align}
	\frac{d}{d t} \frac{\partial L_c}{\partial \dot{\mathbf n}_i} - \frac{\partial L_c}{\partial {\mathbf n}_i}= 0.
	\label{EL1}
\end{align}
The computation yields,
\begin{align}
	I\frac{d^2\mathbf n_i}{d t^2}  = \mathbf E_i + \lambda_i \mathbf n_i.
	\label{EL2a}
\end{align}
Multiplying both sides of this equation by $\mathbf n_i$, we find
\begin{align}
	\lambda_i =-I\bigg(\frac{d\mathbf n_i}{d t}\bigg)^2- \mathbf n_i \cdot \mathbf E_i .
\end{align}
 By substituting $\lambda_i $ into (\ref{EL2a}), we obtain Eq. (\ref{EQM1}).

Using the local spherical coordinates $(\Theta_i,\Phi_i)$ to define the orientation of the dipole, 
\begin{align}
 \mathbf n_i = (\sin\Theta_i \cos\Phi_i, \sin\Theta_i\sin\Phi_i, \cos\Theta_i),
 \end{align}
 one can recast the Euler-Lagrange equations of motion as follows: 
 \begin{align}\label{EL3a}
	\frac{d}{d t} \frac{\partial L}{\partial \dot{\Theta}_i} - \frac{\partial L}{\partial {\Theta }_i}= 0, \\
	\frac{d}{d t} \frac{\partial L}{\partial \dot{\Phi}_i} - \frac{\partial L}{\partial {\Phi}_i}= 0,
	\label{EL3b}
\end{align}
 where $L =  T-U$, and  the kinetic energy of the system is:
 \begin{align}
	T  = \Sigma _i \frac{I}{2 } (\dot\Theta_i^2 + \sin^2 \Theta_i \,\dot \Phi_i^2).
\end{align}

\begin{figure}[tbh]
 \begin{center}
\scalebox{0.2}{\includegraphics{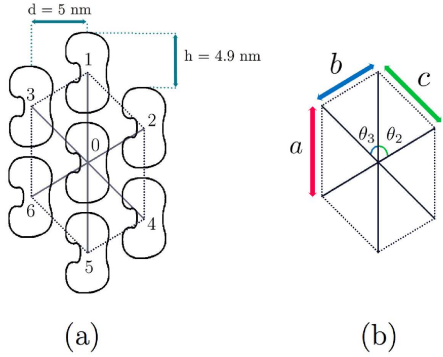}}
 \end{center}
\vspace{-7mm}
 \caption{(Color online) Tubulin neighborhood in the hexagonal unit cell of the microtubule.  The distance between dimers is $d$. The heterodimer helix direction is defined 
 by the height, $h$. The typical values of parameters are: $a=8\, \rm nm$, $b=5.87\, \rm nm$, $c=7.02\,\rm nm$, $d = 5\, \rm nm$, $h = 4.9\, \rm nm$,  $\theta_1 =0$,  $\theta_2 =58.2^{\,\rm o}$, $\theta_3 = 45.58^{\,\rm o}$ \cite{Slyadnikov1,TBCC,THST,GTJT} .
 \label{MT1b}}
  \end{figure}

It is commonly accepted that coupling constants, $J_{ij}$, are nonzero only for the nearest-neighbor dipole moments.  The system of MT dimers may be represented on a triangular lattice, as shown in Fig. \ref{MT1b}, so that each spin has six neighbors.  We denote the constants of interaction between the central dipole in Fig. \ref{MT1b} and nearest neighbors as, $J_{0\alpha}$, and  the distance between the central spin and its nearest neighbors as, $d_{\alpha}$ ($\alpha =1,2, \dots, 6$), setting $d_{01}= d_{04} =a$, $d_{02}= d_{05} =b$, $d_{03}= d_{06} =c$. The corresponding angles (between the central dimer and others) are denoted as, $\theta_1$, $\theta_2$ and $\theta_3$, so that: $ {\mathbf e}_{01}\cdot   {\mathbf e}_{01} =\cos\theta_1$, $ {\mathbf e}_{01}\cdot   {\mathbf e}_{02} =\cos\theta_2$, $ {\mathbf e}_{01}\cdot   {\mathbf e}_{06} =\cos\theta_3$. 

{\em Parameters of the MT.} - The typical values of parameters known from the literature are: $a=8\, \rm nm$, $b=5.87\, \rm nm$, $c=7.02\,\rm nm$, $\theta_1 =0$,  $\theta_2 =58.2^{\,\rm o}$, $\theta_3 = 45.58^{\,\rm o}$ \cite{Slyadnikov1,THST} (See Fig. \ref{MT1b}b.) The radius of the MT can be estimated as, $R \approx 11.2 \,\rm nm$ \cite{TBCC,GTJT}. The unit cell shown in Fig. \ref{MT1b} consists of the central spin surrounded by six neighbors. Its area is: $\Sigma_0= 3ad= 120\, \rm nm^2$.  

To estimate the moment of inertia of a dipole we use the formula for the moment of inertia of thick cylinder: $I= M l^2/12$, where $M$ is the mass of the dipole, and  $l$ is its  length. In our simulations, we take data known from the literature. Assuming: $M \approx (10^{-23} \div  10^{-22} )\,\rm g$ and $l \approx 2 \rm nm$ \cite{Sataric1,Sataric4}, we have: $I\approx 3\dot(10^{-38}\div 10^{-37})\rm g\cdot cm^2$. Using these data, we estimate the parameter J (see Eq.(\ref{HT})) as follows: $J \approx 1.45\cdot 10^{-13}\, \rm erg $.

\subsection{Continuum approximation}

A key question in the studyng of the MT's dynamics is a possibility of use a continuum approximation.
 Recently has been shown  \cite{Zdravkovic14} that for the non-linear model introduced in \cite{Sataric1}  the results obtained in the continuum approximation are in an excellent agreement with the results of the corresponding discrete model. The  findings show that MT can be treated as the continuum system.
 
 The continuum  limit of  the model, described by the Lagrangian, $L= T-U$,  is obtained by allowing the area per a site, $\Sigma_0$, tends to zero, so  that the total area, $N \Sigma_0$, is kept fixed. In this limit, the summation  is replaced by the integral over the MT surface: $ \sum_{\langle ij \rangle} 
\rightarrow (1/2)\int_\Sigma  d^2 x$. The variable, $\mathbf n_i= 
 \mathbf n(\mathbf r_i)$,  should be replaced by a smooth function of the 
 continuum coordinates: $ \mathbf n(\mathbf r_i)\rightarrow  {\mathbf 
 n}(\mathbf r)$. 
  
We find that in the continuum limit the potential energy of the system 
(\ref{H2a}) becomes,
\begin{align}
	&{U} =\frac{J}{\Sigma_0}\int_{\Sigma} \sqrt{g }\,d\Omega \Big (  
	-\frac{\Sigma_0}{2}g^{ij}G_{ab} \partial_i n^a \partial_j n^b  
	\nonumber \\
	&+ G_{ab} n^a n^b + g_0(n^3)^2  +g_1( n^3)^4 - g_2 n^1 \Big ), \nonumber\\
	&i,j =1,2, 	
	\label{HT}
		\end{align}
	where $\Sigma_0= 3ad$ is the area of the unit cell presented in Fig. \ref{MT1b},  $J= 2S^2 \sum^3_{\alpha=1} J_{0\alpha } $, $g_0= PS^2/J$,
$g_1 =QS^4/J$ and $g_2=  BS/J$. 

The local basis is chosen as follows:  $ {\mathbf e}_1=   {\mathbf e}_r$, $  {\mathbf e}_2=   {\mathbf e}_\varphi$, and $ {\mathbf e}_3  ={\mathbf e}_z$, so that one has the following decomposition: $ \mathbf n = n^a{\mathbf e}_a$.
In the cylindrical coordinates the metric on $\Sigma$ can be written as,
\begin{align}
	ds^2 = R^2 d \varphi \otimes d\varphi + d z\otimes d z,
\end{align}
where $R$ is the radius of the MT. In what follows, we use the abbreviation: $\nabla n^a \cdot\nabla n^b = g^{ij}\partial_i n^a \partial_j n^b $.

The metric in the intrinsic space of pseudo-spins is given by:  $G_{ab} =\delta_{ab} - h_{ab}$ $(a,b =1,2,3)$, where
\begin{align}
h_{22} =&\frac{6S^2}{J}\sum^3_{\alpha=1}  J_{0\alpha } \sin^2\theta_\alpha, \\
	h_{23} =&\frac{3S^2}{J} \sum^3_{\alpha=1}  (-1)^\alpha J_{0\alpha } \sin\theta_\alpha \cos\theta_\alpha , \\ 
h_{33} =&\frac{6S^2}{J} \sum^3_{\alpha=1} J_{0\alpha } \cos^2\theta_\alpha   .
\end{align}
 The computation of the constants yields: $h_{22} =1.55$, $h_{23} =0.11$,  $h_{33} =1.45$.
 
Further, it is convenient to introduce the dimensionless coordinates, $\zeta = z/\sqrt {\Sigma_0}$ and $\tilde R = R/\sqrt {\Sigma_0}$. Now, the total action yielding the equations of motion can be written as,
\begin{align}
	S_{tot}=  J  \int  dt  \int_\Sigma  {\mathcal L }d\Sigma + S_\lambda,
		\label{Scl2}
\end{align}
where $d\Sigma = \tilde R d\zeta d\varphi$  and
\begin{align}
	 S_\lambda  = J \int dt\int_\Sigma  \lambda (\mathbf n \cdot \mathbf n-1)\,d\Sigma .
\end{align}
The Lagrangian of the system is given by,
\begin{align}
	{\mathcal L} =\frac{\rho}{2} \bigg(\frac{{\partial\mathbf n} }{\partial t}\bigg )^2 + \frac{1}{2}G_{ab} \nabla n^a \cdot\nabla n^b  - {\mathcal V}(\mathbf n) ,
	\label{L1}
\end{align}
where $\rho = I/J$ and
\begin{align}
{\mathcal V}(\mathbf n) =  G_{ab} n^a n^b + g_0(n^3)^2  +g_1( n^3)^4 - g_2 n^1.
\end{align}
As one can see,  in the continuum limit  the electric properties of the MT are 
described by the nonlinear anisotropic $\sigma$-model \cite{EFR,AMT}. The order 
parameter, $\mathbf n$, is the local polarization unit vector specified by a 
point on the sphere, $S^2$.

The equations of motion are obtained from the variational principle, demanding the total action to be stationary: $\delta S_{tot}=0$. The result is:
\begin{align}
\rho\frac{\partial^2\mathbf n}{\partial t^2}  = \frac{\delta\mathcal L}{\delta{\mathbf n}}
+ \lambda \mathbf n,
	\label{EL3}
\end{align}
where
\begin{align}
\lambda =-\rho\bigg(\frac{\partial\mathbf n}{\partial t}\bigg)^2- \mathbf n\cdot  \frac{\delta\mathcal L}{\delta{\mathbf n}},
\end{align}
and 
\begin{align}
 \frac{\delta\mathcal L}{\delta{\mathbf n}} = \frac{\partial\mathcal L}{{\partial \mathbf n}}  -\nabla \bigg( \frac{\partial\mathcal L}{\partial{\nabla \mathbf n}}\bigg ). 
\end{align}

To simplify the Lagrangian, we will make the following approximation (\ref{L1}). Taking into account that $h_{23},|h_{33}- h_{22}| \ll 1$, we neglect by contributions of these terms and keep only terms with $h_{33}$. This approximation transforms (\ref{L1}) into the following Lagrangian,
\begin{align}
&	{\mathcal L} =\frac{\rho}{2 } \bigg(\frac{{\partial\mathbf n} }{\partial 
	t}\bigg )^2 + \frac{1}{2} (\nabla \mathbf n )^2\nonumber \\
	& - \frac{h}{2}(\nabla n^2 \cdot\nabla n^2 + \nabla n^3 \cdot\nabla 
	n^3 
	 )- {\mathcal W}(\mathbf n) ,
	\label{L2}
\end{align}
where $h=h_{33}$ and
\begin{align}
{\mathcal W}(\mathbf n) =  h( n^1)^2 + g_0(n^3)^2  +g_1( n^3)^4 - g_2 n^1.
\end{align}

 Further, we use the local spherical coordinates $(\Theta,\Phi)$ to define the orientation of the dipole:
$\mathbf n = (\sin\Theta \cos\Phi, \sin\Theta\sin\Phi, \cos\Theta )$. 
Then, the Lagrangian of the system can be recast as follows:
\begin{widetext}
 \begin{align}
	\mathcal L = &\frac{\rho}{2 } ((\partial_t\Theta)^2 + \sin^2 \Theta 
	(\partial_t\Phi)^2) + \frac{1}{2}\big(\big (\nabla \Theta\big )^2 +\big 
	(\nabla \Phi\big )^2\big)
	-\frac{h}{2}(\cos \Theta \sin \Phi\nabla \Theta + \sin \Theta \cos 
	\Phi\nabla \Phi)^2\nonumber \\
	& -\frac{h}{2}\sin^2 \Theta (\nabla \Theta)^2  - 
	\mathcal W(\Theta, \Phi),
	\end{align}
	\end{widetext}
where
\begin{align}
&\mathcal W(\Theta, \Phi) =   ( g_0 - h)\cos^2\Theta  + {g_1} \cos^4\Theta \nonumber \\
&- h\sin^2\Theta \sin^2\Phi - g_2\sin\Theta \cos\Phi .
\end{align}
The Euler-Lagrange equations are
 \begin{align}\label{MA2a}
	\frac{d}{d t} \frac{\partial \mathcal L}{\partial \partial_t{\Theta}} - \frac{\delta \mathcal L}{\delta {\Theta }}= 0, \\
	\frac{d}{d t} \frac{\partial \mathcal L}{\partial \partial_t{\Phi}} - \frac{\delta \mathcal L}{\delta {\Phi}}= 0. 
	\label{MA2b}
\end{align}
One can rewrite these equations as,
\begin{align}\label{A4a}
\rho\frac{\partial^2 \Theta}{\partial t^2} = \frac{\delta \mathcal L}{\delta {\Theta }}, \\
	\rho\frac{\partial}{\partial t}\bigg(\sin^2\Theta\frac{\partial \Phi}{\partial t} \bigg )  = \frac{\delta \mathcal L}{\delta {\Phi }}.
	\label{A4b}
\end{align}

 \subsection{Ground state}

\begin{figure}[tbh]
 \begin{tabular}{c}
 	\scalebox{0.19}{\includegraphics{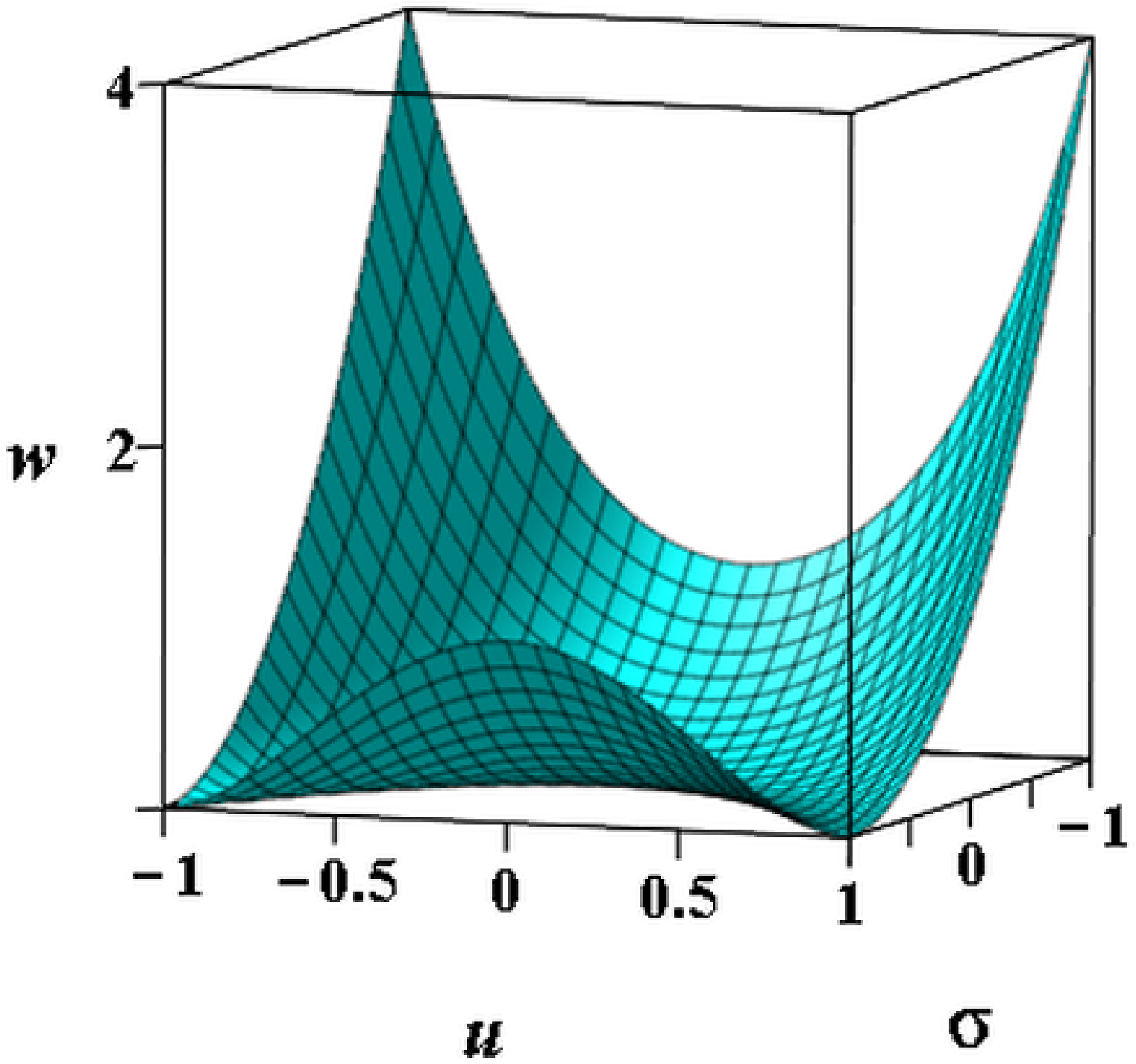}}\\
 (a)
  \end{tabular}
   \begin{tabular}{c}
 \scalebox{0.19}{\includegraphics{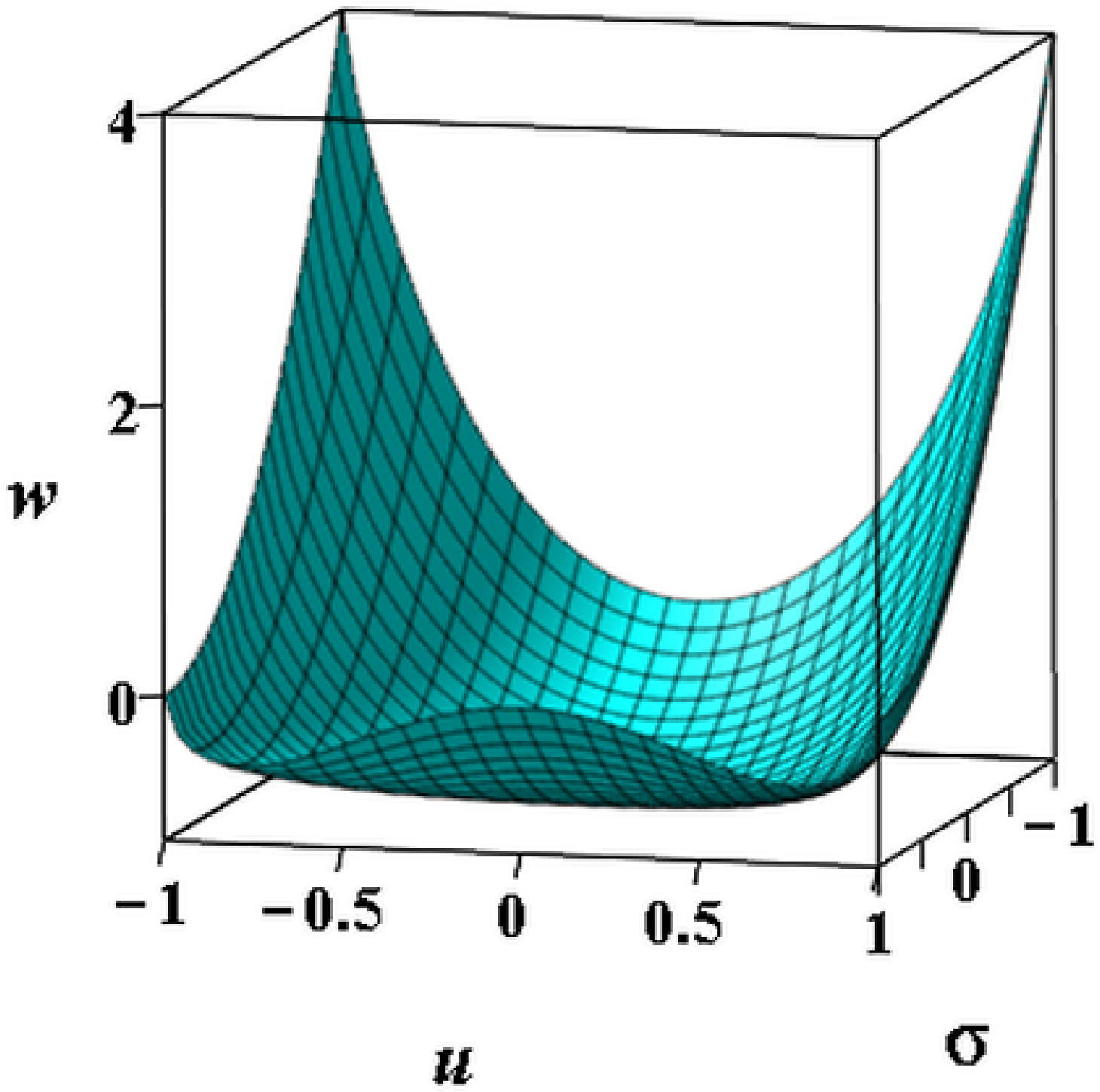}}\\
 (b)
  \end{tabular}
  \begin{tabular}{c}
 	\scalebox{0.19}{\includegraphics{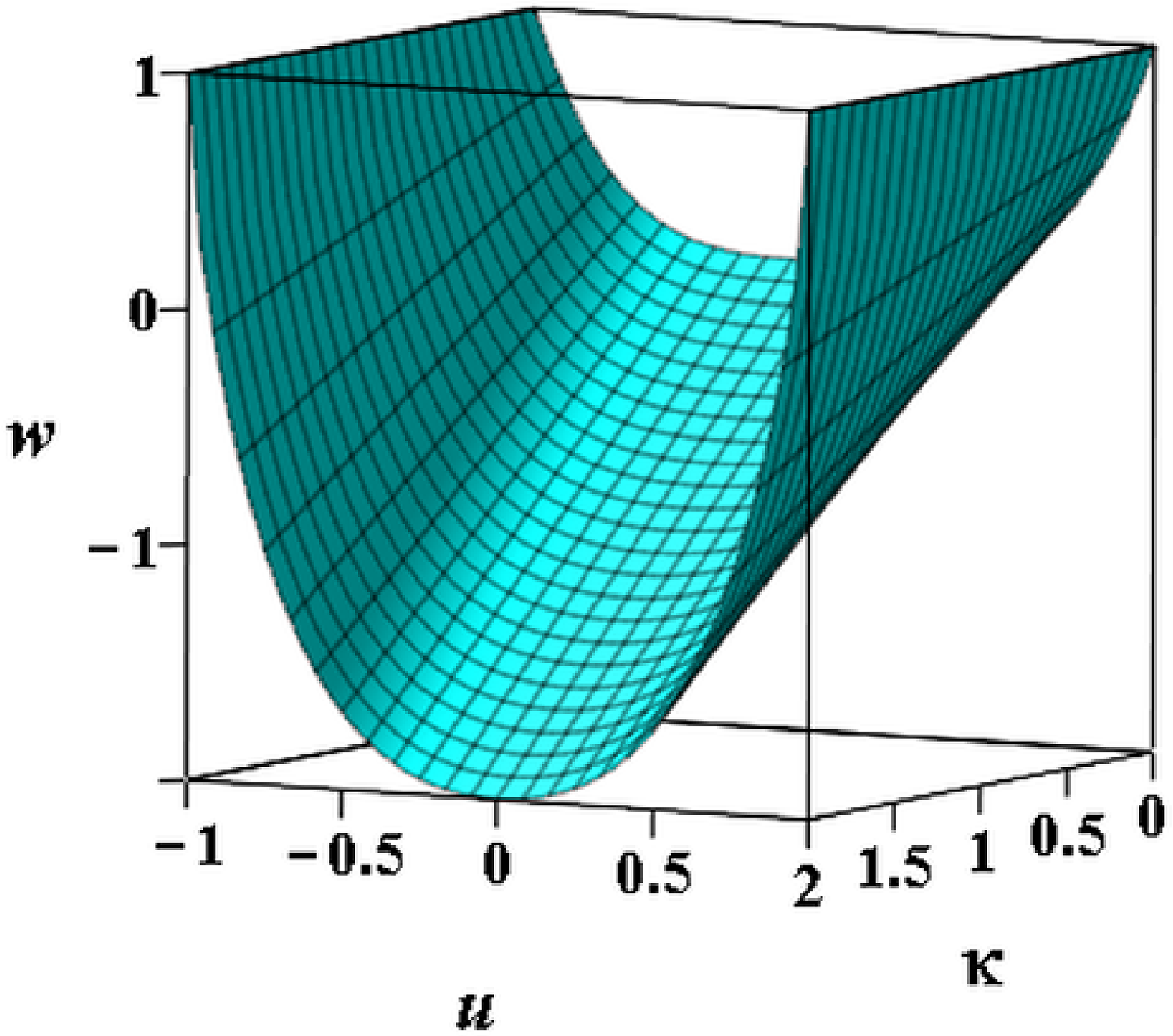}}\\
 (c)
  \end{tabular}
   \begin{tabular}{c}
 \scalebox{0.19}{\includegraphics{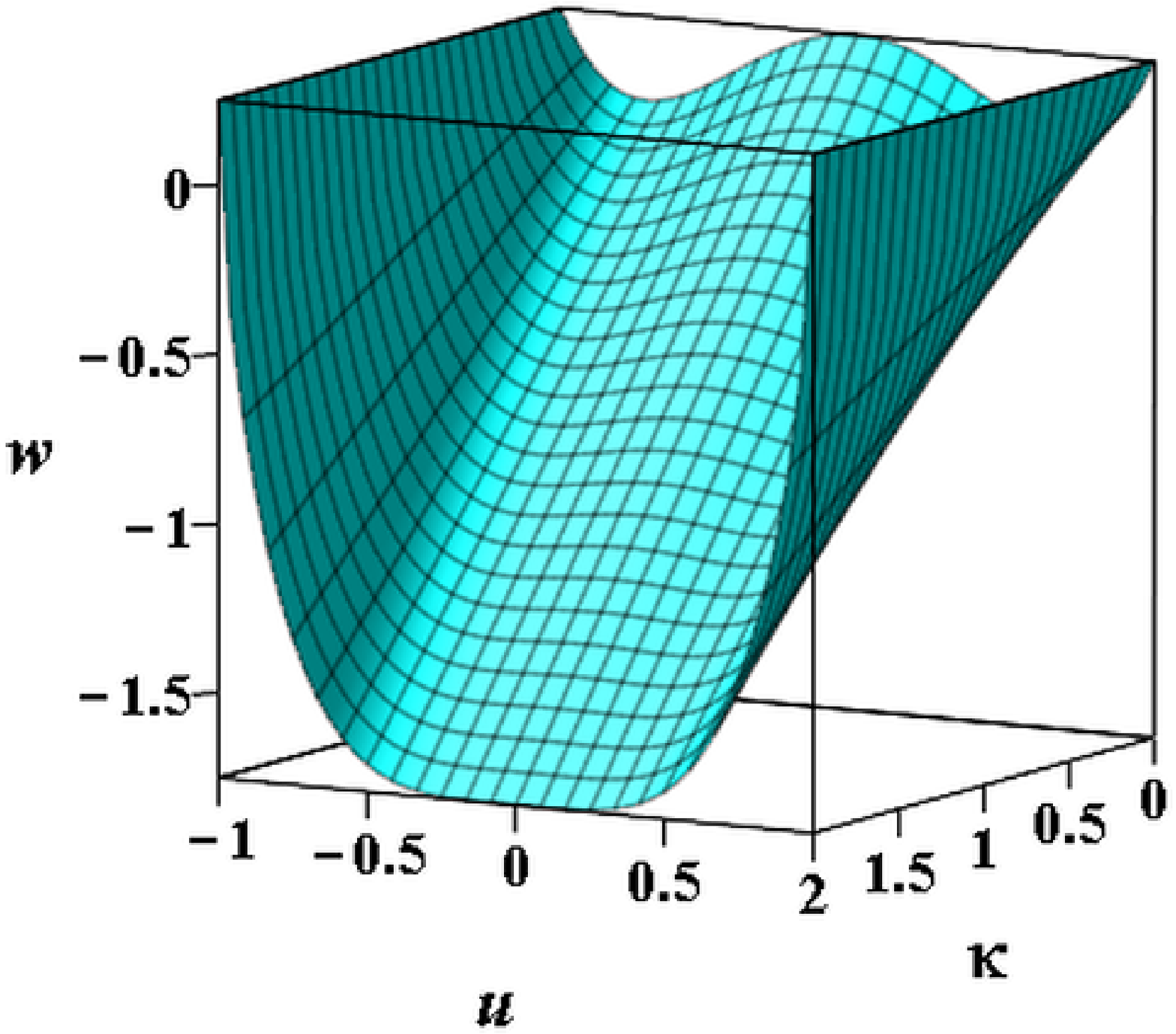}}\\
 (d)
  \end{tabular}
 \caption{ Dimensionless  energy  density $w$ as a function of $u$ and parameters $\sigma$ and $\kappa$. (a,b) $w$ vs $u$ and $\sigma$: (a) $\kappa=0$, (b) $\kappa=1$; (c,d) $w$ vs $u$ and $\kappa$: (c) $\sigma=0$, (d) $\sigma=0.5$.
 \label{En1}}
  \end{figure}

The ground state of the MT, yielding the permanent dipole moment with $\Phi =0$, is defined by the minimum value of  the energy,
\begin{align}
	 E (u) =E_0 +  J \int_\Sigma {\mathcal V}(u) \,d\Sigma	,
	 \label{W1}
	\end{align}
where  $u = \cos \Theta$, 
\begin{align}
	E_0 &=- J g_1 \int_\Sigma  \sigma^2 \,d\Sigma ,
	\label{ED}
\end{align}
and 
\begin{align}
 {\mathcal V}= {g}_1\big(   (\sigma -u^2)^2  -\kappa\sqrt{1- u^2}\big) .
 \label{U1}
\end{align}
Here we set $\sigma = (h-g_0)/(2g_1)$ and $\kappa= g_2/g_1$.  One can see that there are three critical points: $u_1=0$, and $u_{2,3}$ defined from the equation:
\begin{align}
  u^6 -(1+2\sigma)u^4 +\sigma(2+\sigma )u^2 +\kappa^2/16- \sigma^2=0.
 \label{P3}
\end{align}
The behavior of the dimensionless energy density of the system, $ w = {\mathcal V}/{g}_1$, as a function of $u$ and parameters $\sigma$ and $\kappa$ is presented in Fig. \ref{En1}.

First, we consider the case when the parameter $\kappa=0$.  In this case, the critical points of the Hamiltonian are given by
 \begin{align}
 u_1 &=0, \\
u_{2,3}& = 
\pm\sqrt{\sigma}.
\label{IM2a}
\end{align}

As one can see, if  $\sigma<0$, the ground state  of the MT is paraelectric, 
$u_1=0$. It corresponds to the  radial orientation of the permanent dipole moments of the tubulin dimers with respect to the surface of the MT (Fig. \ref{MT}). 
For $\sigma >0$, the homogeneous ground state is a doubly degenerate ferroelectric state. The dipole momentum of the tubulin dimer is given by $u_{2,3}=\pm\sqrt{\sigma}$ (see Fig. \ref{En1}a).

\begin{figure}[tbh]
 \begin{center}
  \scalebox{0.25}{\includegraphics{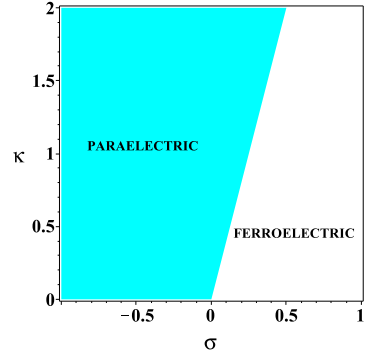}}
  \end{center}
\vspace{-8mm}
 \caption{The phase diagram.
 \label{D1}}
  \end{figure}

As it follows from the phase diagram presented in Fig. \ref{D1}, when  $\kappa >4\sigma$, the ground state  of the MT is paraelectric. It corresponds to the  radial orientation of the permanent dipole moments of the tubulin dimers with respect to the surface of the MT.  When $\kappa <4\sigma$, the ground state of the system is ferroelectric. 
  
  Note, that the consideration of the ground state is done here at zero temperature. The finite temperature effects are discussed, for example,  in \cite{TBHM}. In particular, it is argued in \cite{TBHM}, that the critical temperature of the order-disorder transition depends on the values of the dipole moment and on the electric permittivity.

\section{Nonlinear dynamics in the continuum limit}
  
 In order to construct a solution for a nonlinear wave moving along the MT with the constant velocity, we use the traveling wave ansatz. We assume that, in the cylindrical coordinates, the  field variables are functions of  
 \begin{align}
 	 \xi = \sqrt{\frac{2}{\eta p\Sigma_0}}(z + h_0 \varphi/2\pi -v t),
 \end{align}
 where  $\eta = h/g_1$ and $p=1+(h_0/2\pi R)^2$, the velocity of the wave being $v$. Then, one can show that the field equations (\ref{EL3}) possess the first integral of motion: 
\begin{align}
	\frac{\rho}{2 } \bigg(\frac{{\partial\mathbf n} }{\partial t}\bigg )^2 + \frac{1}{2}G_{ab} \nabla n^a \cdot\nabla n^b  + {\mathcal V}(\mathbf n)= \rm const.
	\label{IM1b}
\end{align}
 
\begin{figure}[tbh]
 \begin{center}
  \scalebox{0.3}{\includegraphics{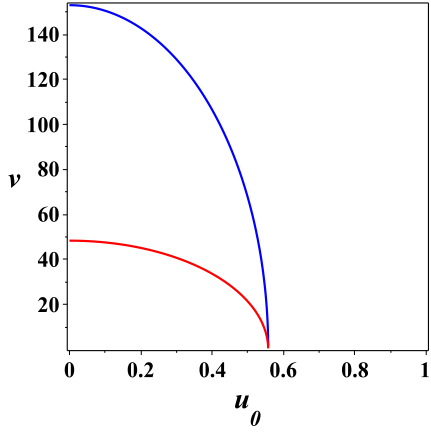}}
  \end{center}
\vspace{-8mm}
 \caption{Velocity of the excitation ($m/s$): $M =10^{-23}  \,\rm g$ (blue), $M =10^{-22}  \,\rm g$ (red), $l = 2\, \rm nm$.
  \label{V1}}
  \end{figure}

For the Lagrangian (\ref{L2}) we obtain,
\begin{widetext}
	\begin{align}
\frac{\rho}{2} \bigg(\frac{{\partial\mathbf n} }{\partial t}\bigg )^2 + \frac{1}{2} (\nabla \mathbf n )^2 - \frac{h}{2}(\nabla n^2 \cdot\nabla n^2 + \nabla n^3 \cdot\nabla n^3 )+ {\mathcal W}(\mathbf n) =\rm const.
	\label{TW1}
\end{align}
\end{widetext}

In the local spherical coordinates $(\Theta,\Phi)$,  Eq. (\ref{TW1}) can be rewritten as,
\begin{widetext}
	\begin{align}
	& (u_0^2 - \cos^2 \Theta  ) \bigg( \frac{d \Theta}{d \xi} \bigg)^2 + \sin^2 \Theta   \Big (u_0^2 - \frac{1}{h} \cot \Theta -\sin^2 \Phi \Big ) \bigg( \frac{d \Phi}{d \xi} \bigg)^2 + \frac{1}{2}\sin (2\Theta) \sin (2\Phi) \frac{d \Theta}{d \xi} \frac{d \Phi}{d \xi} \nonumber\\
	&  - (\sigma - \cos^2\Theta)^2 +  \eta\sin^2\Theta \sin^2\Phi  + \kappa\sin\Theta \cos\Phi	= \rm const,
	 \label {TW1a}
	  \end{align}
\end{widetext}
where  $u_0^2 =  1- 1/ h-\rho v^2/(hp \Sigma_0)$. This yield a simple formula for the nonlinear wave propagation velocity 
\begin{align}
	v=\sqrt{(\sigma_0^2-u_0^2 )\frac{hp\Sigma_0 }{\rho}},
\end{align}
where we set $\sigma_0^2 = 1 -1/h$.

In Fig. \ref{V1}, the dependence of the velocity of the wave on the parameter $u_0$ is depicted. We find that the velocity of the wave is limited: $v \leq v_0$, where $v_0 \approx 155 \rm m/s$.

\subsection{Particular solutions: $\Phi =0$}

 Employing (\ref{TW1a}), we will seek a solution of Eqs. (\ref{MA2a}) - (\ref{MA2b})  in the form: $\Phi =0$ and $\Theta =\Theta(\xi)$. One can show that $\Phi =0$ satisfies  Eq. (\ref{MA2b}), and for the function, $\Theta(\xi)$, we obtain the nonlinear differential equation,
 \begin{align}
& (u_0^2 - \cos^2\Theta) \frac{d^2 \Theta}{d \xi^2} +  \frac{1 }{2} \sin(2\Theta)\bigg( \frac{d \Theta}{d \xi} \bigg)^2 \nonumber \\
& -\sin(2\Theta) (\sigma - \cos^2\Theta) + \frac{\kappa }{2}\cos \Theta =0.
\label{Theta}
 \end{align}

The qualitative properties of the system one can elucidate by applying the standard technique for studying of the dynamical systems by means of the phase space \cite{Zasl}. To depict the phase portrait of the system we use Eq. (\ref{TW1a}) written  as,
\begin{align}
&(u_0^2 -\cos^2\Theta)\Big (\frac{d \Theta}{d\xi}\Big)^2 
 -   (\sigma -\cos^2\Theta)^2  \nonumber \\
& +\kappa\sin\Theta= \rm const .
 \label{MA8}
\end{align}
\begin{figure}[tbh]
  \begin{tabular}{c}
 	\scalebox{0.25}{\includegraphics{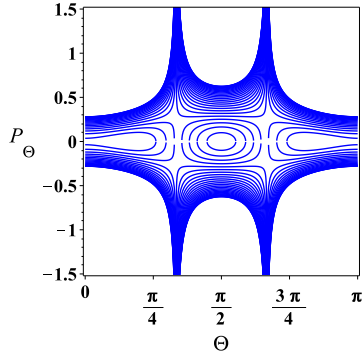}}\\
 (a)
  \end{tabular}
  \begin{tabular}{c}
 	\scalebox{0.25}{\includegraphics{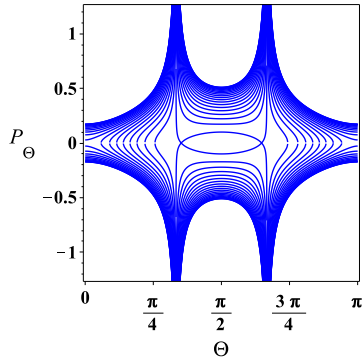}}\\
 (b)
  \end{tabular}
   \begin{tabular}{c}
 	\scalebox{0.25}{\includegraphics{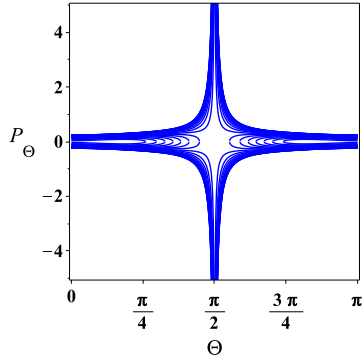}}\\
 (c)
  \end{tabular}
 \caption{ Phase portrait of the system (\ref{MA8}). The momentum 
 $P_\Theta$ is defined as,  $P_\Theta = d\Theta/d\xi$.  (a) $\sigma = 
 0.75$, $u_0 =0.5$; (b) $\sigma = 0.2$, $u_0 =0.5$; (c) $\sigma = 0.75$, 
 $u_0 =0$. In all cases: $\kappa=0$
 \label{PP1}}
  \end{figure}

\begin{figure}[tbh]
    \begin{tabular}{c}
 \scalebox{0.25}{\includegraphics{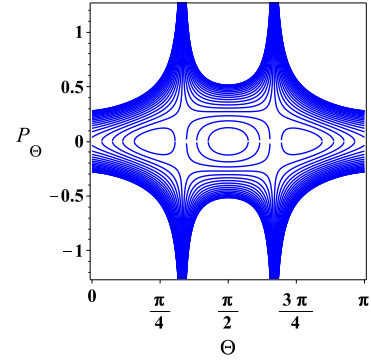}}\\
 (a)
  \end{tabular}
  \begin{tabular}{c}
 	\scalebox{0.25}{\includegraphics{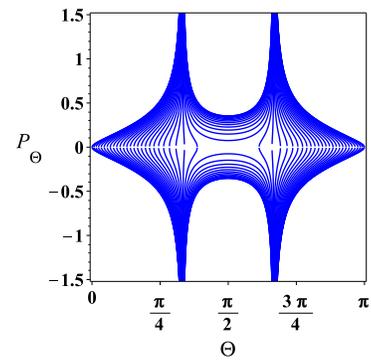}}\\
 (b)
  \end{tabular}
   \begin{tabular}{c}
 \scalebox{0.25}{\includegraphics{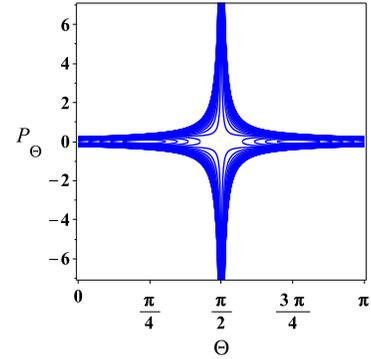}}\\
 (c)
  \end{tabular}
 \caption{ Phase portrait of the system (\ref{MA8}). The momentum 
 $P_\Theta$ is defined as,  $P_\Theta = d\Theta/d\xi$.  (a) $\sigma = 
 0.75$, $u_0 =0.5$; (b) $\sigma=0$, $u_0 =0.5$, ; (c) $\sigma=0.75$,  
 $u_0 =0$.  In all cases: $\kappa=0.5$.
 \label{PP1a}}
  \end{figure}

In Figs. \ref{PP1} and \ref{PP1a},  the phase portraits of the system (\ref{MA8}) are demonstrated in the plane ($\Theta, P_\Theta $), for different parameters, where $P_\Theta = d\Theta/d\xi$. One can observe the occurrence of the three elliptic points for $\sigma > u_0^2$ (Fig. \ref{PP1}a). When $\sigma < u_0^2$, two elliptic points disappear.

\begin{figure}[tbh]
\begin{tabular}{c}
 \scalebox{0.5}{\includegraphics{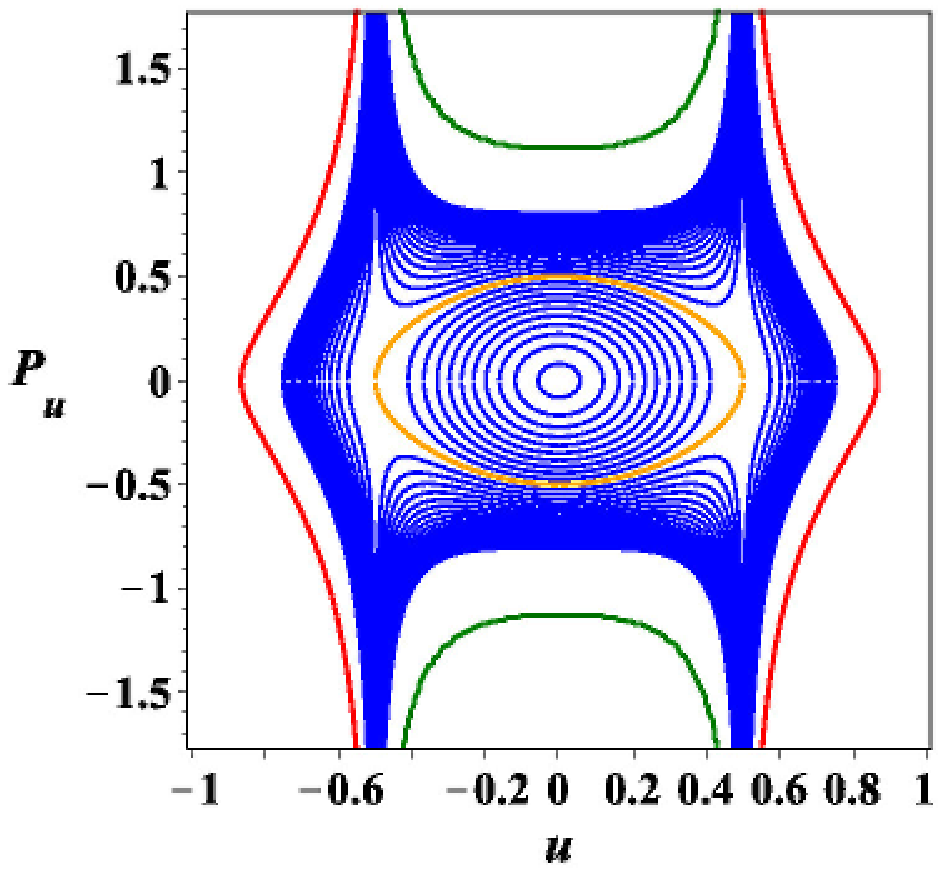}}\\
 (a)
  \end{tabular}
  \begin{tabular}{c}
 	\scalebox{0.5}{\includegraphics{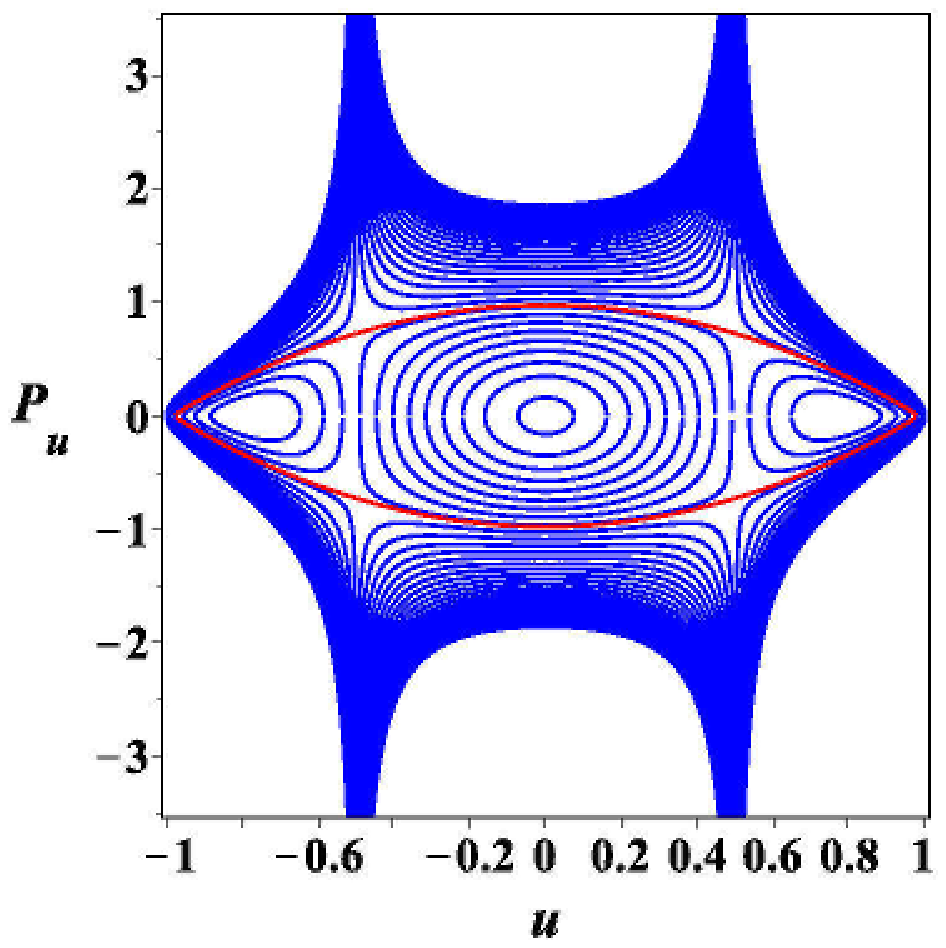}}\\
 (b)
  \end{tabular}
 \caption{ Phase portrait of the system (\ref{MA8a}) in the plane ($u,P_u$): (a) $\sigma =u_0^2 = 0.25$, $k=0.5$; (b) $\sigma = 0.6$, $k=0.975$. Parameters: $u_0 =0.5$, $\kappa =0$.
 \label{PP2}}
  \end{figure}

By substitution $u = \cos \Theta$ into Eq. (\ref{MA8}), one can rewrite it  as,
\begin{align}
&\frac{u_0^2 -u^2}{ (1 - u^2)} \Big (\frac{d u}{d\xi}\Big)^2 d 
 -   (\sigma -u^2)^2 \nonumber \\
 & +\kappa\sqrt{1- u^2}= \rm const.
 \label{MA8a}
\end{align}
Denoting the constant of integration as, $-\varepsilon$, one can rewrite this equation as,
\begin{align}
\bigg( \frac{d u}{d \xi} \bigg)^2	+ V(u) =0,
\label{U2}
\end{align}
where 
\begin{align}
	V(u) =- \frac{( (\sigma -u^2)^2 - \kappa\sqrt{1- u^2} - \varepsilon)(1-u^2) }{u_0^2 -u^2}.
\end{align}
Thus, the dynamics of the dipoles on the surface of the MT can be considered as the motion of the effective particle of mass $m= 2$ in the potential $V(u)$, with the total energy of the system being, $E=0$. In Fig. \ref{PP2}, the phase portrait of the system (\ref{U2}) is shown in the plane ($\Theta, P_u $), where $P_u = du/d\xi$. 

\subsubsection{Snoidal waves and kinks: $\kappa =0$}

\begin{figure}[tbh]
 \scalebox{0.36}{\includegraphics{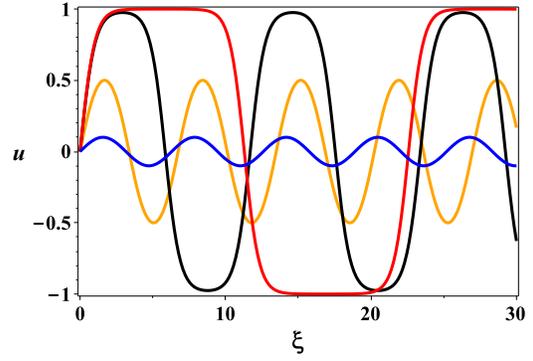}}
 \caption{The  sn-solution: $k = 0.1$ (blue), $k = 0.5$ (orange),  $k = 0.975$ (black), $k = 0.9999$ (red)
 \label{Sn2a}}
  \end{figure}

Here we assume $\kappa=0$, that implies absence of the intrinsic radial electric field ($g_2 =0$). 
Choosing the constant of integration in  Eq. (\ref{MA8})  as, 
$\varepsilon = (\sigma - u_0^2)^2$, we obtain,
\begin{align}
\bigg( \frac{d u}{d \xi} \bigg)^2	={( 2\sigma -u_0^2-u^2)(1-u^2)}.
\label{M1}
\end{align}
 Assume $u_0^2<2\sigma < 1+u_0^2$, then the analytical solution of this equation is given by a snoidal wave,
\begin{align}
u =k \,{\rm sn}(\xi -\xi_0,k).
 \label{J1}
\end{align}
Here $k= \sqrt{2\sigma -u_0^2}$, and ${\rm sn}(z,k)$ is the Jacobi elliptic function. In Fig. \ref{Sn2a} the sn-solutions for different choices of the constant $k$ are depicted. In Fig. \ref{PP2}a, the orbit for $k=0.5$ is presented by the orange curve.

The period of the sn-wave is given by $T= 4K$, where 
\begin{align}
	K = \int_0^{\pi/2}\frac{d 
	\varphi}{\sqrt{1- k^2 \sin^2 \varphi}},
\end{align}
is the complete elliptic integral of the first kind \cite{abramo}. 

For $k^2 \ll 1 $ and $k'^2 = 1- k^2 \ll 1$, applying the Maclaurin Series in $k^2$ and $k'^2$ \cite{abramo}, we obtain 
\begin{align}
	&u=k \sin\xi -\frac{k^3}{4}(\xi - \sin\xi\cos \xi )\cos \xi + 
	{\mathcal O}(k^5),\\
	&u= \tanh\xi -\frac{k'^2}{4}(\xi +\sinh\xi\cosh \xi ){\rm sech}^2 \xi 
	+ {\mathcal O}(k'^4).
\end{align}
(For simplicity, here we set $\xi_0 =0$.)

 In particular, for $k=0$, we obtain $u=0$. This solution corresponds to the  elliptic point located at the center of the phase space in Fig. \ref{PP2}.  When $k=1$,  the sn-waves become the kink
\begin{align}
	u = \tanh (\xi - \xi_0),
	\label{K1}
\end{align}
with the following boundary conditions: $u(\pm \infty) =\pm 1$. (See Fig. \ref{Kink}.) In Fig. \ref{PP2}b, the corresponding orbit is presented by separatrix (red curve).
\begin{figure}[tbh]
\begin{tabular}{c}
 \scalebox{0.35}{\includegraphics{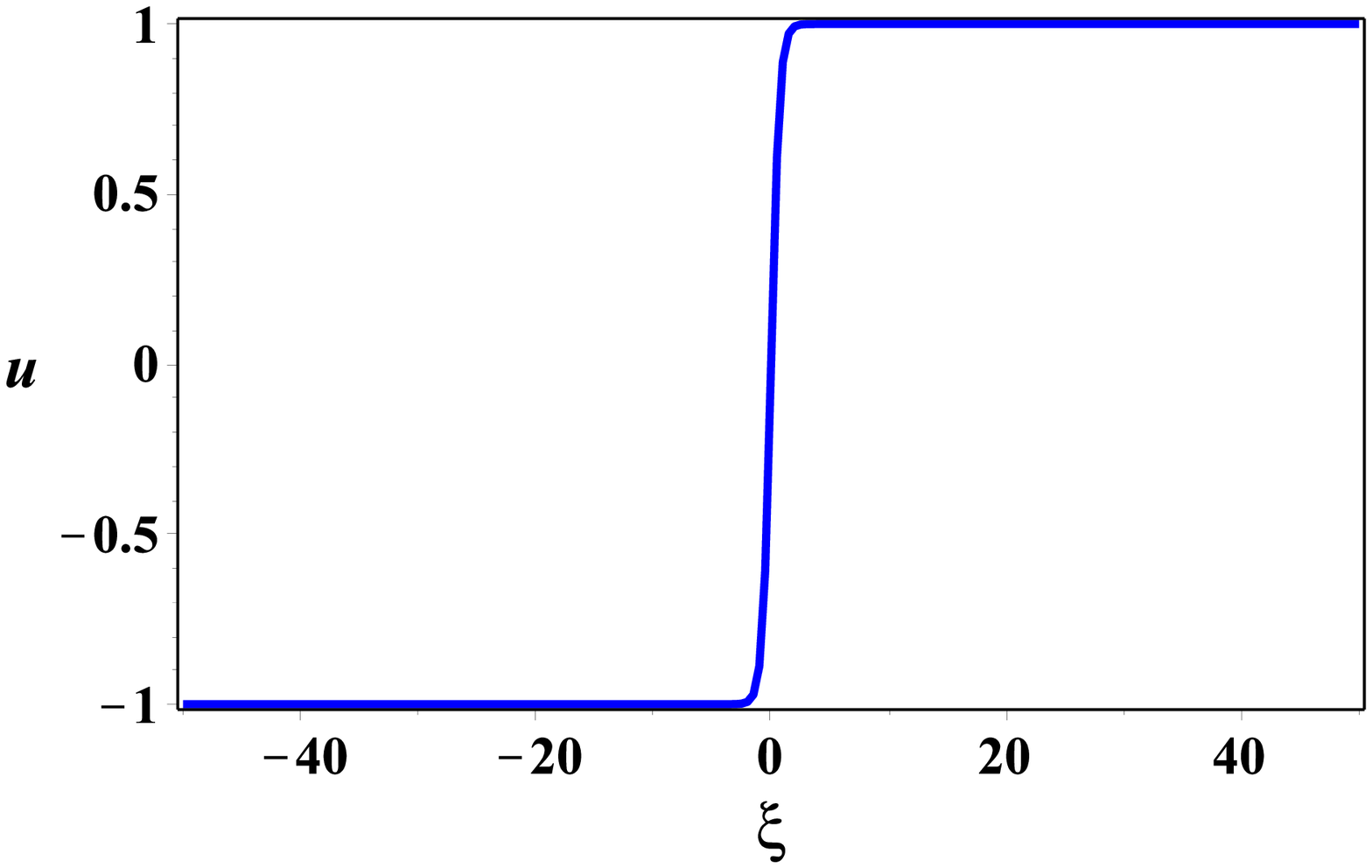}}\\
 (a)
  \end{tabular}
  \begin{tabular}{c}
 	\scalebox{0.35}{\includegraphics{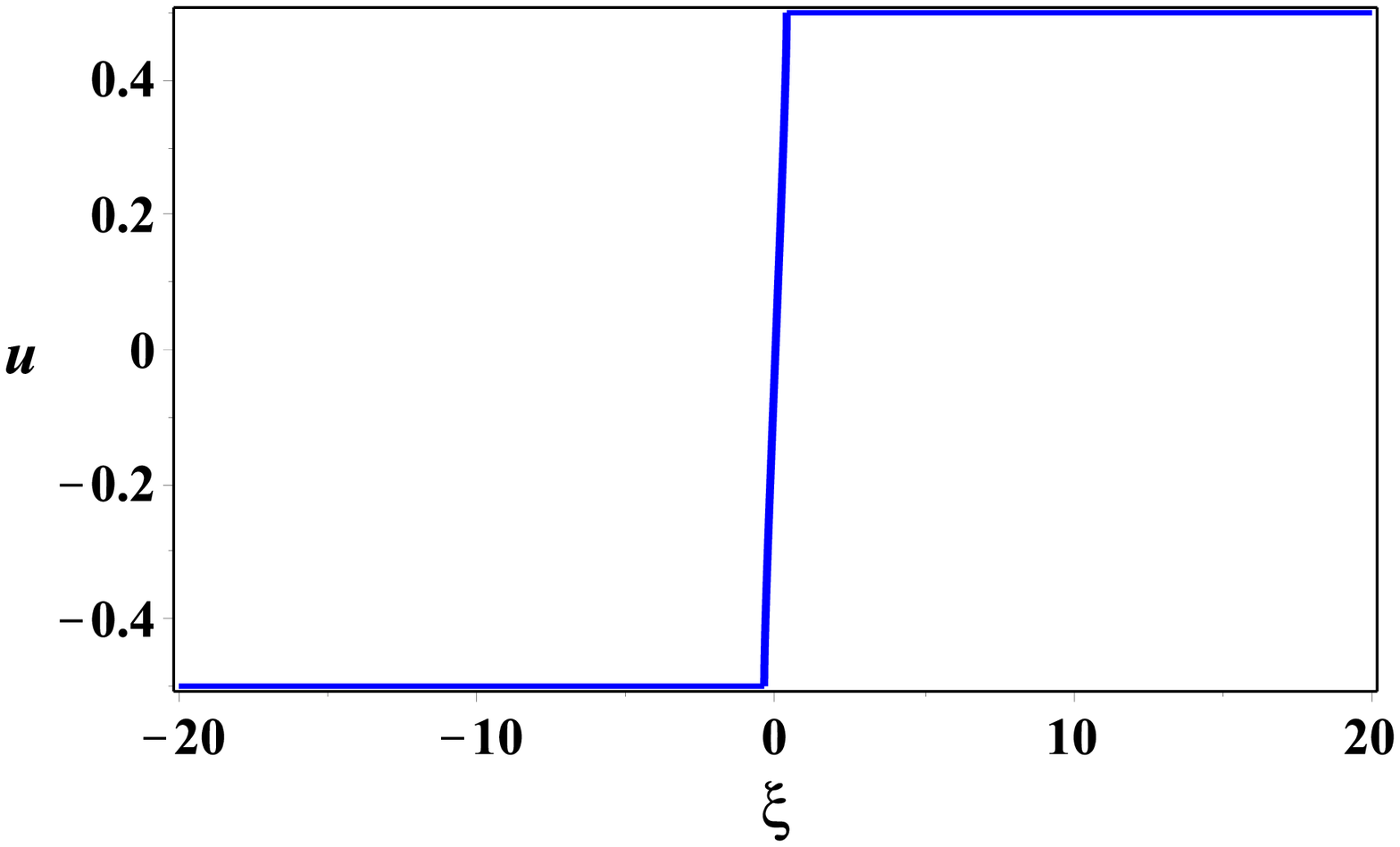}}\\
 (b)
  \end{tabular}
 \caption{Kink. (a) analytical solution,   $\sigma  = 0.25$, $\varepsilon= -0.25$. (b)  numerical solution, $\sigma=0.625$, $\varepsilon = 0.141$. Parameters: $u_0 =0.5$,  $\kappa =0$.
 \label{Kink1}}
  \end{figure}
  
  A topological classification of kinks is given in terms of homotopy group \cite{MND}. The topological charge, $\pi_0$, of kink is determined by the magnitude, $n_z$ of the polarization vector at the ends of the MT:
\begin{align}
	\pi_0 = \frac{1}{2}(n_z(+\infty) -n_z(-\infty)).
\end{align}
To change the topological charge one needs to overcome the potential barrier, proportional to the size of the MT (formally, infinite potential barrier).

 In Fig. \ref{Kink1} the analytical solution (\ref{K1}) is depicted.  In Fig. \ref{Kink1}b, the  numerical kink solution for 
 $\varepsilon = 0.141$ is shown. In Fig. \ref{PP2}b, the corresponding orbit is presented by the green curve.

\subsubsection{Spikes: $\kappa =0$}

A spike solution can be obtained as excitation of the ground state, $u_g$.  To estimate energy carried by  spike, we approximate it by step function. Then, using Eq. (\ref{W1}),  we obtain
\begin{align}
	\Delta w_{sp} =  w_{g} - w_{sp}    =- Jg_1 ( u_{sp}^2  -u_g^2)^2,
\end{align}
where $u_{sp}$ is the height of the spike, and  $w_g = -J g_1  u_g^4$ is the energy density of the ground state (see Eq.(\ref{ED})).

\begin{figure}[tbh]
\begin{center}
 \scalebox{0.3}{\includegraphics{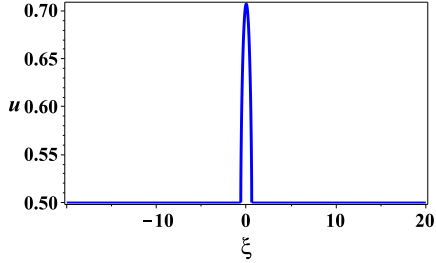}}
   \end{center}
 \caption{ Spike: $\varepsilon= 0.25$, $u_0 =0.5$, $\sigma = 0.25$, $\kappa =0$.
 \label{SP1}}
  \end{figure}
 In the mean field approximation, the electric field in $z$-direction of the MT being in the ground state, can be obtained by using the relation: $w_g =-\mathbf S_g \cdot \mathbf E$.  Let us assume that all dipoles are aligned along the MT, that implies $u_g =1$. Then,  the electric filed due to permanent dipole reaches its maximum magnitude given by
\begin{align}
	E_z^{\rm max }= \frac{J g_1}{S}.
	\label{E1}
\end{align}
Using this result,  one can estimate the electric field produced by the spike as, 
\begin{align}
\Delta E_z = E_z^{\rm max} (u_{sp}^2 - u_g^2)^2.
\end{align}
The maximum value of the electric field produced by spike can be estimated as follows: $\Delta E_z \leq \Delta E_z^{\max} $, where
\begin{align}
	\Delta E_z^{\max} =  E_z^{\rm max} (1 - u_g^2)^2.
		\label{SP2}
\end{align}
Let $\Theta_0$ be the angle between the permanent dipole and axis orthogonal to the surface of the MT. Then, (\ref{SP2}) can be rewritten as
\begin{align}
	\Delta E_z^{\max} = { E_z^{\rm max}} \cos^4\Theta_0 \leq E_z^{\rm max} .
	\label{SP3}
\end{align}
Thus, the maximum magnitude of the electric field produced by spike is bounded by $E_z^{\rm max}$. 
 
As it is discussed in the literature, in the ground state the orientation of the dipoles with respect to the surface of the MT can be defined by $\Theta_0 \approx 29^{\,\rm o}$  \cite{TBCC}. Substituting these data into Eq. (\ref {SP3}), we obtain the following estimation for the electric field produced by the spike: $\Delta E_z^{\max} \approx  0.6 { E_z^{\rm max}} $. To evaluate $E_z^{\rm max}$, we use  data available for the electric field inside of  the MT: $E_z \sim 10^5 \div 10^8 \,\, \rm V/m$ \cite{Sataric1}. Then, we obtain the following estimate for the electric field produced by the spike: 
\begin{align}
\Delta E_z^{\rm max} \lesssim & \,0.6\cdot (10^5 \div 10^8 )\,\, \rm V/m \nonumber \\
&=(0.06 \div 60 )\,\, \rm mV/nm .
	\end{align}

In Fig. \ref{SP1}, the localized spike solution is presented. In the phase space in Fig. \ref{PP2} the corresponding orbit is indicated by the red curve on the right. 

Note, that both the soliton and spike solutions could be important for information and signal transduction, given that they both may transfer information in a dissipation-free way.  

\subsection{Particular solutions: $\Theta =\pi/2$}

\subsubsection{Chiral solitons}

In this section, we study solution related to the paraelectric ground state. We seek a solution of Eqs. (\ref{MA2a}) - (\ref{MA2b})  in the form: $\Theta =\pi/2$. One can show that $\Theta =\pi/2$ satisfies  Eq. (\ref{MA2a}).  Substituting $\Theta =\pi/2$  into Eq.  (\ref{TW1a}), we obtain
\begin{align}
	&   \Big (u_0^2  -\sin^2 \Phi \Big ) \bigg( \frac{d \Phi}{d \xi} \bigg)^2  \nonumber\\
	& +  \eta \sin^2\Phi  + \kappa \cos\Phi	= \rm const,
	 \label{CS}
\end{align}

Introducing a new function, $u_\varphi = \sin \Phi$, one can recast this equation as,
\begin{align}
\ \Big (\frac{d u_\varphi}{d\xi}\Big)^2 
+ U ( u_\varphi)  = 0,
 \label{CS1a}
\end{align}
where
\begin{align}
	U(u_\varphi) = \frac{\big( \varepsilon - \eta u_\varphi^2 - \kappa\sqrt{1 -u_\varphi^2} \big)(1 - u_\varphi^2)}{ u_\varphi^2 -u_0^2}.
\end{align}

We denote by $\varepsilon$ the constant of integration in Eq. (\ref{CS}).
\begin{figure}[tbh]
\begin{tabular}{c}
 \scalebox{0.55}{\includegraphics{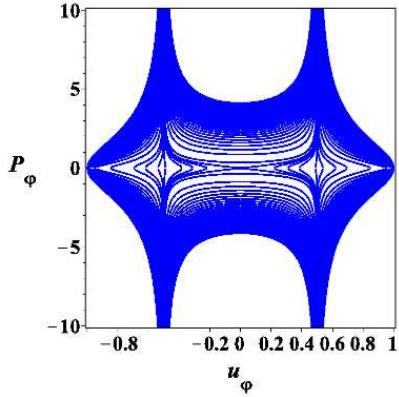}}\\
 (a)
  \end{tabular}
  \begin{tabular}{c}
 	\scalebox{0.55}{\includegraphics{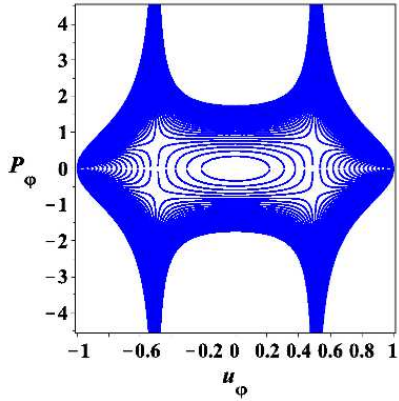}}\\
 (b)
  \end{tabular}
 \caption{Phase portrait of the system (\ref{CS1a}) in the plane ($u_\varphi,P_\varphi$): (a) $\eta = 0.1$, $\kappa=0.75$; (b) $\eta = 0.75$, $\kappa=0.25$. Parameters: $u_0 =0.5$.}
 \label{CS1}
   \end{figure}
\begin{figure}[tbh]
\scalebox{0.65}{\includegraphics{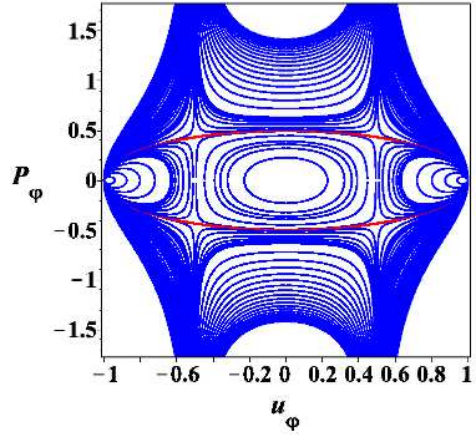}}\\
 \caption{ Phase portrait of the system (\ref{CS1a}) in the plane ($u_\varphi,P_\varphi$):  $\eta = 0.25$, $\kappa=0$, $u_0 =0.5$.} 
 \label{CS3}
   \end{figure}

A chiral solutions correspond a boundary conditions:
\begin{align}
	\cos \Phi|_{}\pm \infty = \frac{\kappa}{2\eta} \pm \sqrt{\frac{\kappa^2}{4\eta^2}-\varepsilon}.
\end{align} 
A chirality is a topological charge, being described  by the relative homotopy group \cite{MND}, and defined as follows:
\begin{align}
	\chi= \frac{1}{\pi} \int_{-\infty}^{\infty} dz\,{\mathbf e}_z\cdot(\mathbf n \times\bigg(\Big( \frac{\partial \mathbf n}{\partial z}\Big)\bigg).
\end{align}
Chiral solitons can produce quantized charge transport across the MT that is topologically protected and controllable by the soliton's chirality.

Employing the spherical coordinates, one can recast this equation as follows:
\begin{align}
	\chi= \frac{1}{\pi} \int_{-\infty}^{\infty} dz\, \sin^2 \Theta \frac{\partial \Phi}{\partial z}.
	\end{align}
Taking into account that in our case $\Theta =\pi/2$, we obtain
\begin{align}
	\chi= \frac{1}{\pi} (\Phi(+\infty) - \Phi(-\infty)).
	\end{align}
Chiral solitons in the phase space are presented by orbits located in the interval $(-u_0,u_0)$. (See Figs. \ref{CS1} and \ref{CS3}.) 

Suppose that $\kappa =0$, then taking the constant of integration as, $\varepsilon = \eta u_0^2$, one can rewrite (\ref{CS}) as:
\begin{align}
\Big (\frac{d u_\varphi}{d\xi}\Big)^2 
= \eta{(1 - u_\varphi^2)}.
 \label{CS2}
\end{align}
The analytical solution of this equation is given by 
\begin{align}
	u_\varphi = \sin (\sqrt{\eta}(\xi - \xi_0)).
\end{align}
The corresponding orbit is presented in Fig. \ref{CS3} by separatrix (red curve).

\subsection{Two-dimensional representation of solutions}

The solutions obtained in the previous sections have the form: $\Theta = \Theta(z+ \nu \varphi - vt)$ and
 $\Phi= \Phi(z+ \nu \varphi - vt)$.  Thus, they describe the two-dimensional nonlinear waves propagated on the surface of the MT, along the $z$-direction.  
 
\begin{figure}[tbh]
  \begin{tabular}{c}
 	\scalebox{0.325}{\includegraphics{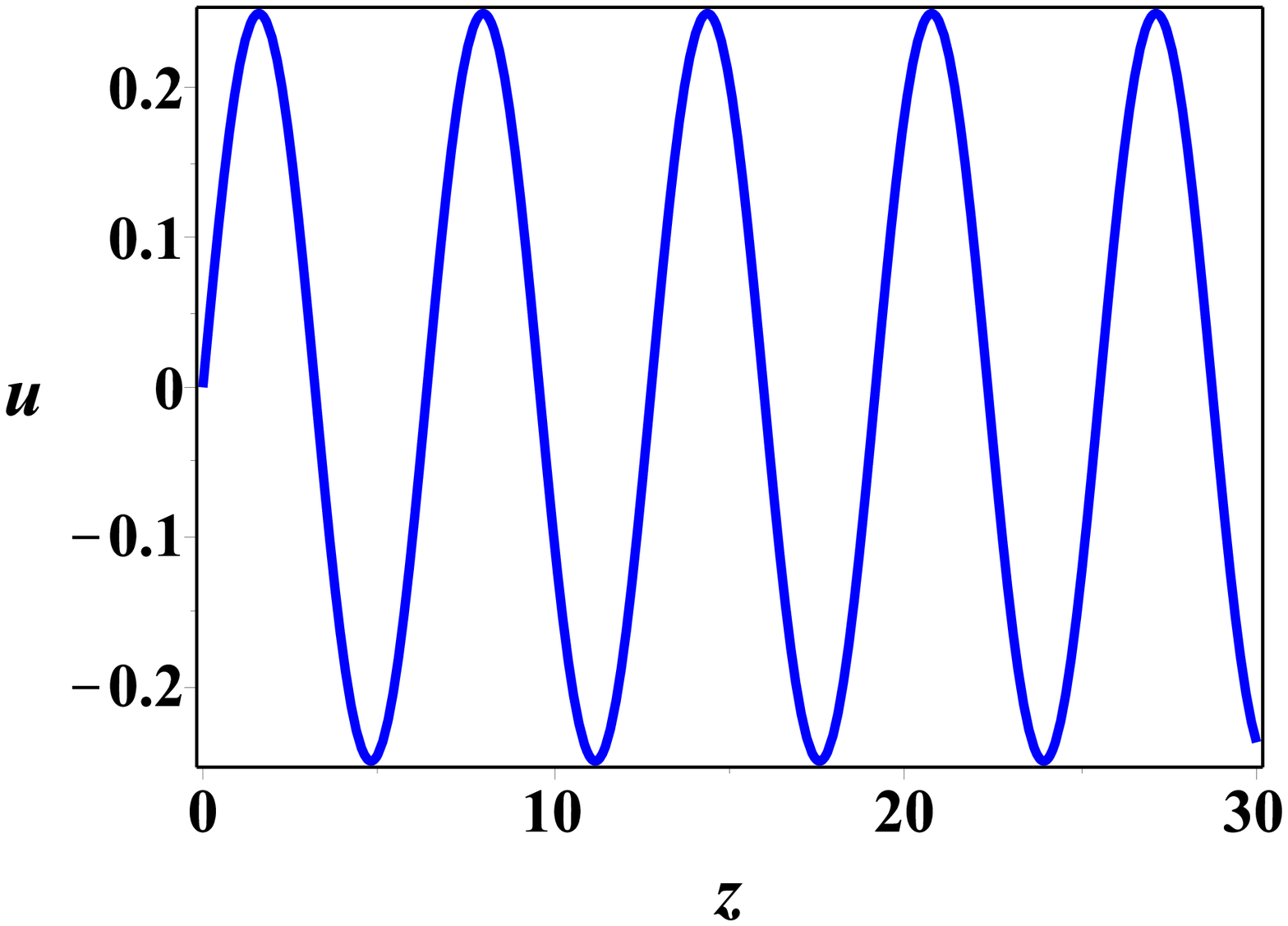}}\\
 (a)
  \end{tabular}
  \begin{tabular}{c}
 	\scalebox{0.375}{\includegraphics{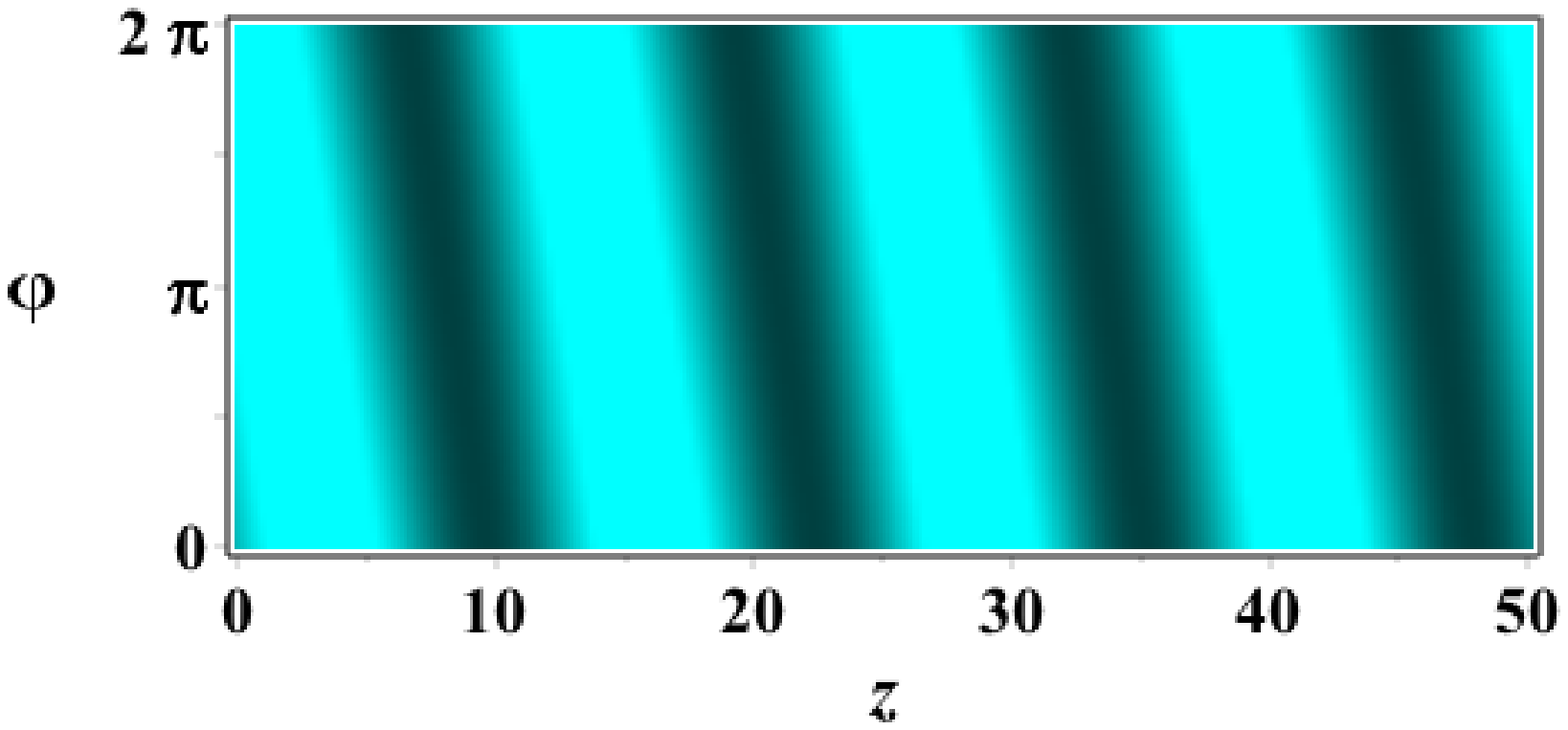}}\\
 (b)
  \end{tabular}
   \begin{tabular}{c}
 	\scalebox{0.375}{\includegraphics{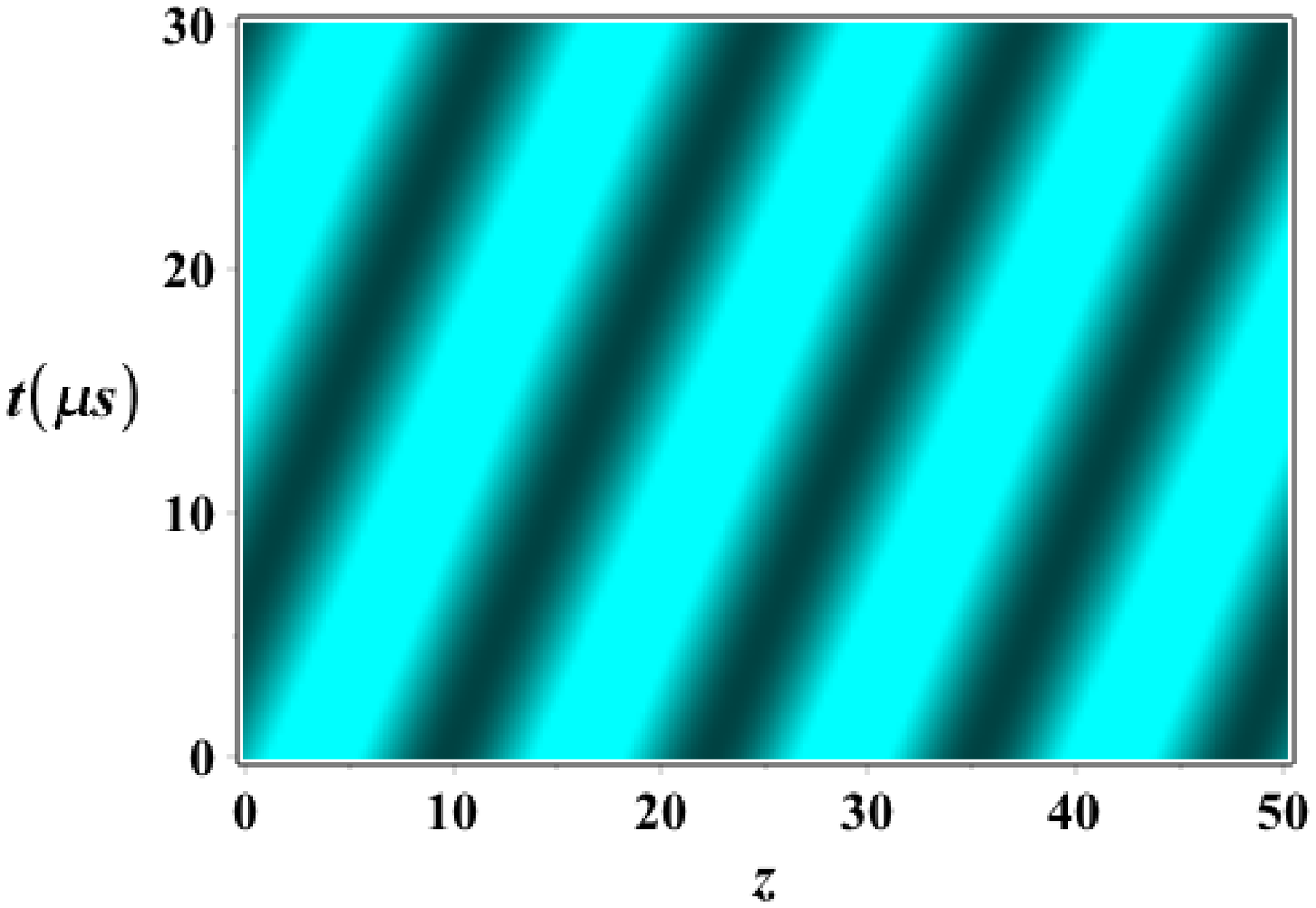}}\\
 (c)
  \end{tabular}
 \caption{Sn-solutions. (a) $u$ vs $z$ ($v=\nu =0$); (b) Density plot  of the helicoidal static snoidal solution $v=0$. Density plot of the propagating sn-wave along the MT ($\varphi = \rm conts$). Parameters: $v= 0.1 \rm m/s$,  $\nu= 100 \,\rm  nm$, $C= 0.5$, $k=0.25$.
 \label{sn}}
  \end{figure}

\begin{figure}[tbh]
  \begin{center}
 \scalebox{0.35}{\includegraphics{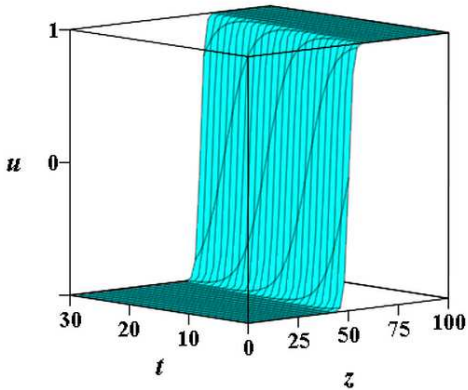}}
 \end{center}
 \caption{Propagating kink excitation. Parameters: $v= 0.1 m/s$,  $\nu= 100 \,\rm  nm$, $C= 0.5$, $k=0.25$.
 \label{Kink}}
  \end{figure}

 In Fig. \ref{sn}a,b,  the static  helicoidal sn-solution  is depicted.  In Fig.  \ref{sn}c, the helicoidal sn-wave is presented. In Fig. \ref{Kink}, the  solution, describing kink moving in the $z$-direction, is depicted. All parameters are given in the corresponding figure captions.
 \begin{align}
&(u_0^2 -\cos^2\Theta)\Big (\frac{d \Theta}{d\xi}\Big)^2 
 -   (\sigma -\cos^2\Theta)^2  \nonumber \\
& +\kappa\sin\Theta= \rm const .
 \label{MA9}
\end{align}

\section{Discussion and conclusion}
In this paper, we introduced and studied theoretically a generalized pseudo-spin model for describing the nonlinear static and dynamic solutions in the tubulin-protein microtubule.
The ``pseudo-spin" means that the length of the dipole for each turbulin-based heterodimer is constant. 
The main advantage of our model is that it includes such relevant effects as: geometry of the heterodimers positioned on the cylindrical surface of the microtubule; realistic dipole-dipole interactions; an external electric field produced by the solvent; the additional electric potential responsible for possible degeneracy of the dipole energy at each heterodimer. Staring from a discrete model of interacting dipoles, we reduce our consideration to the continuum approximation, which results in nonlinear partial differential equations for the pseudo-spin. Note, that these equations are different from the well-known Bloch equations for the average spin.  

The partial solutions of these equations include snoidal waves, solitons, kinks, and localized spikes. These solutions have specific structures, and they can be useful for a better understanding of many effects  associated with the functional properties of microtubules. In particular, the obtained spike solutions can serve as good candidates for static and dynamic memory bits and for electric excitations responsible for information transfer processes.

Experimental verification of the results obtained in this paper will represent a significant interest.

Before closing we would like to make some remarks on the comparison of our solutions with previously studied solitons in MT. From a mathematical point of view such solitons have also appeared in simplified conformal chain models of MTs considered in \cite{Nick11,Nick12,Nick13}, where however the relevant degree of freedom was the projection of the displacement vector of a dimer along the $z$-axis of the MT, in the context of simple ferroelectric-ferrodistortive lattice models of MTs~\cite{Sataric1}, upon taking the continuum limit. In these models interactions among the spin chains is also modeled by a double-well potential of the displacement vector in simplest cases, although more general models, leading to more complicated solitonic states have been proposed in \cite{Nick11,Nick12,Nick13}. The current model, using the pseudo spin approach, appears to take better account of realistic geometrical and physiological features than the above conformal spin chain models. 

 The classical solitonic solutions we have found can be modified by quantum corrections, as in the models considered in \cite{Nick11,Nick12,Nick13,Nick21,Nick22}. There are standard WKB techniques that provide such modifications, which may turn out to be physically important in MT, should quantum effects play a role. In this sense, classical solitonic solutions may be viewed as macroscopic coherent states of a quantum spin system. For such states to exist one needs sufficient isolation of the MT dimer system from external entanglement. We have argued in \cite{Nick11,Nick12,Nick13,Nick21,Nick22} that such an isolation is possible as a result of string dipole-dipole interactions between the ordered water molecules in the interior of the MJT cavities and the neighboring dimer walls. In {\it in vivo} situations such strong interactions may overcome thermal losses and provide the necessary environmental isolation, 
as proposed to happen in the cavity model of MT~\cite{Nick11,Nick12,Nick13}, in which a thin (a few Angstrom think) cavity layer between the MT interior and the dimer wall 
acts like an isolated cavity, leading to relatively long decoherence time (up to microseconds, for moderately (micron long) MT.

The role of ordered water, and other details of the structure of the MT have been ignored in our treatment above. It would be interesting to incorporate them in future studies of these systems. It may well be that once this is done, we can disover more realistic solitonic structures of {\em helical shape} that are responsible for information and signal transduction in a {\em dissipation-free} way.
Moreover, if such quantum effects are at play, there may be long distance correlations between parts of the MT system (`quantum wiring') in analogy with such claimed long lasting (femtoseconds) effects in algae~\cite{algae}, as mentioned previously. Ferroelectricity  might be important for sustaining such effects~\cite{Nick11,Nick12,Nick13,Nick21,Nick22}.

\begin{acknowledgments}  	
   The work by G.P.B. was carried out under the auspices of the National Nuclear Security Administration of the U.S. Department of Energy at Los Alamos National Laboratory under Contract No.
DE-AC52-06NA25396. A.I.N. and M.F.R. acknowledge the
support from the CONACyT.  The work of N.E.M. is partially supported by STFC (UK) under the research grant ST/L000326/1.
\end{acknowledgments}

\end{document}